\documentclass[pre,aps,twocolumn]{revtex4-1}
\usepackage{graphicx}
\usepackage{bm}
\usepackage{tabularx}
\usepackage{threeparttable}

\begin{document}

\title{Eigenstates of Thiophosgene Near the Dissociation Threshold  - Deviations From Ergodicity}

\author{Srihari Keshavamurthy\\
{\em 
Department of Chemistry, Indian Institute
of Technology, Kanpur, Uttar Pradesh 208016, India}
}

\begin{abstract}
A subset of the highly excited eigenstates of thiophosgene (SCCl$_{2}$) near the dissociation threshold are analyzed using sensitive measures of quantum ergodicity. We find several localized eigenstates, suggesting that the intramolecular vibrational energy flow dynamics is nonstatistical even at such high levels of excitations. The results are consistent with recent observations of sharp spectral features in the stimulated emission spectra of SCCl$_{2}$. 
\end{abstract}

\maketitle

\section{Introduction}
\label{intro}

The statistical Rice-Ramsperger-Kassel-Marcus (RRKM)\cite{rrkmorig} theory of reaction rates occupies a central place in the field of reaction dynamics both for its elegance and simplicity. Indeed, within the RRKM approximation reaction rates can be calculated\cite{rrkmbook} irrespective of the intricate intramolecular dynamics that happen prior to the reaction. In recent years, however, the appearance of several examples\cite{nonrrkm1,nonrrkm2,nonrrkm3,nonrrkm4,nonrrkm5,nonrrkm6,nonrrkm7} of intrinsically non-RRKM reactions have rekindled an old question\cite{oxrice,leonberne} - when is a system expected to be in the RRKM regime? From one perspective, answering the question requires understanding the nature of the molecular eigenstates near reaction thresholds. In particular, the situation wherein all the eigenstates are sufficiently delocalized seems ideal for the validity of the statistical approximation. On the other hand, existence of localized eigenstates at the reaction threshold would suggest strong deviations from the RRKM regime. Consequently, there have been many studies\cite{dynassign1,dynassign2,dynassign3,dynassign4,dynassign5,dynassign6,dynassign7,dynassign8,dynassign_ks} aiming to characterize the nature of highly excited eigenstates in different systems. Recent reviews\cite{Kellrev,farantosrev} highlight the relationship between eigenstate assignments based on classical phase space structures and deviations from the RRKM  regime.

The above arguments, more precisely, are connected to a question that has confounded researchers for nearly a century - 
what constitutes a quantum analog of the classical ergodic hypothesis in an isolated many body quantum system?  Progress towards answering this question has mainly come from  semiclassical analysis of quantum chaotic models\cite{qchaosbook}, giving rise to the notions of weak quantum ergodicity\cite{wqe1,wqe2} and quantum unique ergodicity\cite{que}.  Recently, an approach that has attracted substantial attention is the so called eigenstate thermalization hypothesis (ETH)\cite{eth_sred}. According to ETH, for an isolated quantum system governed by the Hamiltonian $\hat{H}$ with eigenstates $|\alpha \rangle$, the eigenstate expectation value $V_{\alpha \alpha} \equiv \langle \alpha |\hat{V}|\alpha \rangle$ of an observable $\hat{V}$ changes slowly and smoothly with the state\cite{eth_sred,eth_deutsch,eth_peresfeingold}. Specifically, 
\begin{equation}
\langle \alpha|\hat{V}| \beta \rangle \equiv V_{\alpha \beta} = {\cal V}(E_{\alpha}) \delta_{\alpha \beta} + O(\hbar^{(f-1)/2})
\label{ethmat}
\end{equation}
with the $O(\hbar^{0})$ leading term of ${\cal V}(E_{\alpha})$ being the classical microcanonical average\cite{wqe1,wqe2}
\begin{equation}
\langle V \rangle_{\rm mc}(E_{\alpha}) = \frac{\int V({\bf q},{\bf p}) \,\delta[H({\bf q},{\bf p})-E_{\alpha}]d{\bf q} d{\bf p}}{\int  \delta[H({\bf q},{\bf p})-E_{\alpha}]d{\bf q} d{\bf p}}
\label{scdvz}
\end{equation}
In the above equation, $f$ is the number of degrees of freedom and $V({\bf q},{\bf p})$ is the classical symbol corresponding to the quantum $\hat{V}$. Furthermore, consider the long-time average of the time-dependent expectation value $\langle j(t)|\hat{V}| j(t) \rangle \equiv \langle V(t) \rangle$ 
\begin{equation}
\bar{V} \equiv \lim_{t\rightarrow \infty} \langle V(t) \rangle = \sum_{\alpha} |C_{j\alpha}|^{2} V_{\alpha \alpha}
\end{equation}
in some initial nonstationary state $|j \rangle = \sum_{\alpha} C_{j\alpha} |\alpha \rangle$ with mean energy $\bar{E}_{j}$.  Since ETH implies that $V_{\alpha \alpha}$ is approximately constant in an appropriately chosen energy window of width $\Delta E$ one can write
\begin{eqnarray}
\bar{V} &=& \langle V \rangle_{\rm mc}(E_{j}) \nonumber \\
&\equiv &\frac{1}{N_{\Delta E}} \sum_{\alpha}^{|E_{j}-E_{\alpha}| < \Delta E} V_{\alpha \alpha}
\label{ethcor}
\end{eqnarray}
with $N_{\Delta E}$ being the number of eigenstates within the energy window.
These predictions of ETH, along with the effect of finite state space, have been tested in a variety of systems like interacting hard core bosons on lattices\cite{ethbec1,ethbec2} and one-dimensional interacting spin chains\cite{ethspinchains}. 

The issue of whether all classically nonintegrable quantum systems obey ETH is still under debate\cite{eth_debate}. 
In particular, many body Anderson localization can lead to systems showing both thermalized and localized phases\cite{aleth}. Recent studies\cite{ethhom} show that ETH can still hold as long as the various perturbations act homogeneously {\it i.e.,} do not have any specific selection rules, leading to ergodicity over the entire relevant Hilbert space. Interestingly, the phenomenon of  intramolecular energy flow, which is at the heart of RRKM, has intimate connections\cite{lwjcp,ksacp} to Anderson localization in terms of the state space model\cite{ivrsp}.  In the state space model two different classes of initial states called as the edge and interior have different mechanisms for exploring the quantum state space with the dynamics of the former class being dominated by dynamical tunneling\cite{leitwoly96}. Moreover, the dynamics of the initial states as determined by the various anharmonic resonances is typically inhomogeneous\cite{parksjcp},  suggesting that ETH might be generically violated in effective Hamiltonian models describing intramolecular energy flow.
In an early work\cite{leitwoly96} Leitner and Wolynes provided a criterion for quantum ergodicity and have argued that facile energy flow throughout the state space is possible if the interior states are extended. Thus, assuming quantum effects such as dynamical tunneling\cite{davhel,stumarcus,ksdyntun} can efficiently couple\cite{leitwoly96} the edge states to the interior states, one expects that violation of ETH for the interior states signals the system being in the non-RRKM regime.

Clearly, as evident from Eq.~\ref{ethmat} and Eq.~\ref{ethcor}, the central quantities of concern are the various eigenstate expectation values $V_{\alpha \alpha}$. There is also a close connection between the $V_{\alpha \alpha}$ and the parametric evolution of energy levels with changing system parameters, also known as level-velocities\cite{pechukas,gaspard}. The literature on level-velocities and curvatures as tools to characterize quantum chaos is quite extensive and we refer to only a select few studies here\cite{lvelrefs}. Three very early studies, however, are worth mentioning. The first one is a study\cite{weissjort} of the highly excited states of the Henon-Heiles and Barbanis Hamiltonians using the level-velocity approach by Weissman and Jortner. The second study\cite{ramkay} on the local ergodicity probes in Henon-Heiles system by Ramachandran and Kay in fact is an early example of ETH. The third study\cite{nordhrice} by Nordholm and Rice puts forward a definition of quantum ergodicity which also is very close to the spirit of ETH and it is interesting to compare the arguments leading up to their definition with those provided in the more recent work\cite{ethbec2} by Rigol and Srednicki.
Later studies have used the eigenstate expectation values to identify special classes of localized eigenstates in systems with several degrees of freedom.  For instance, eigenstate expectation values have been utilized before to dynamically assign the highly excited states of several systems\cite{dynassign_ks}. Very recently\cite{hiller_tiltbec}, using the technique of parametric level motion, Hiller et al. have identified robust states of ultracold bosons in tilted optical lattices coexisting with extensively delocalized states. Existence of such robust few-body bosonic states and dynamically assignable states in molecules are indicative of a lack of statisticality in the system. As an example, in a recent work Leitner and Gruebele have shown\cite{leitgrueb} that significant corrections to the RRKM rates for conformational reactions can be associated with different classes of eigenstates that exist near the reaction threshold. In the molecular context, the eigenstate viewpoint is naturally related to the nature of the intramolecular vibrational energy redistribution (IVR) dynamics at the energies of interest. Thus, Jacobson and Field have used the time-dependent expectation values of the anharmonic resonances to understand the origins of unusually simple stretch-bend dynamics in acetylene\cite{jacfield}.

In this work the ETH perspective is utilized to investigate the nature of the eigenstates of thiophosgene (SCCl$_{2}$) near its dissociation energy ($\approx600$ THz). The motivation for this study comes from the recent work\cite{chowgrueb1} by Chowdary and Gruebele wherein they observed sharp assignable features near and above the dissociation energy in the stimulated emission pumping spectra. Previous works\cite{jungtay,thioprev1,thioprev2} have focused on the nature of the eigenstate and IVR dynamics near the threshold for onset of IVR ($\approx240$ THz) and have already established the existence of different classes of localized eigenstates and hence nonstatistical IVR dynamics. Employing a model spectroscopic Hamiltonian it is shown that localized states do persist persist at energies close to the dissociation energy. Moreover,  it is shown that there are special initial states that are robust despite the increased density of states and strong anharmonic couplings. 

\section{Model Hamiltonian}

The Hamiltonian used for the current study is a fairly accurate effective Hamiltonian for SCCl$_{2}$
constructed\cite{thioprev1} by Sibert and Gruebele using canonical Van Vleck perturbation theory and can be expressed as $H = H_{0} + V_{\rm res}$ with
\begin{eqnarray}
H_{0} & = &\sum_{i} \omega_{i} \left(v_{i} + \frac{1}{2} \right) + 
             \sum_{ij} x_{ij} \left(v_{i} + \frac{1}{2} \right) \left(v_{j} + \frac{1}{2} \right) \nonumber \\
        &+& \sum_{ijk} x_{ijk} \left(v_{i} + \frac{1}{2} \right) \left(v_{j} + \frac{1}{2} \right) \left(v_{k} + \frac{1}{2} \right) 
\end{eqnarray}
being the zeroth-order anharmonic Hamiltonian and the various anharmonic resonances, coupling the zeroth-order states, are contained in
\begin{eqnarray}
V_{\rm res} &=& k_{526} a_{2}^{\dagger} a_{5} a_{6}^{\dagger} + k_{156} a_{1} a_{5}^{\dagger} a_{6}^{\dagger} \nonumber \\
                   &+& k_{125} a_{1}^{\dagger} a_{2}^{\dagger} a_{5}^{2} + k_{36} a_{3}^{2} a_{6}^{\dagger 2}  \nonumber \\
                   &+& k_{261} a_{1}^{\dagger}a_{2}a_{6}^{2} + k_{231} a_{1}^{\dagger} a_{2} a_{3}^{2} + {\rm c.c}
\label{qham}
\end{eqnarray}
 In the above expression, the operators $a_{k}$ and $a_{k}^{\dagger}$ are
lowering and raising operators for the $k^{\rm th}$ mode respectively with $[a_{k},a_{k}^{\dagger}]=1$, in analogy to the
usual harmonic oscillator operators. Note that there are other much weaker resonances in the Hamiltonian which are ignored in the current study. The above Hamiltonian has effectively three degrees of freedom due to the existence of three, approximately, conserved quantities or polyads
\begin{eqnarray}
K &=& v_{1} + v_{2} + v_{3}  \nonumber \\
L &=& 2v_{1} + v_{3} + v_{5} + v_{6}  \nonumber \\
M &=& v_{4}
\label{polyads}
\end{eqnarray}
Therefore $H$ is block diagonal and in the present work the focus is on $(K,L,M)=(13,25,14)$ block which has a total of $1365$ eigenstates spanning an energy range of about $(18099,20841)$ cm$^{-1}$. Note that Chowdary and Gruebele used the above Hamiltonian to perform an extensive study of the vibrational state space and concluded\cite{chowgrueb2} that about $1$ in $10^{3}$ zeroth-order states are localized near the dissociation energy. In what follows, for convenience, we will refer to the resonances by the highlighting the modes involved. Thus, the first term in Eq.~\ref{qham} will be referred to as the $526$-resonance.

The classical Hamiltonian corresponding to Eq.~\ref{qham} can be expressed\cite{jungtay} in terms of action-angle variables as $H({\bf J},{\bm \psi})=H_{0}({\bf J})+V_{\rm res}({\bf J},{\bm \psi})$ where
\begin{eqnarray}
H_{0}({\bf J},{\bm \psi}) &=& C + \sum_{i=1,3} \varpi_{i} J_{i} + \sum_{ij} \alpha_{ij} J_{i}J_{j} \nonumber \\
                                      &+& \sum_{ijk} \beta_{ijk} J_{i} J_{j} J_{k}
\end{eqnarray}
is the zeroth-order Hamiltonian and the resonant perturbations given by
\begin{eqnarray}
V_{\rm res}({\bf J},{\bm \psi}) &=& v_{156}({\bf J};K_{c},L_{c}) \cos \psi_{1} \nonumber \\
                                              &+& v_{526}({\bf J};K_{c},L_{c}) \cos \psi_{2} \nonumber \\
                                              &+& v_{125}({\bf J};K_{c},L_{c})\cos(\psi_{1}+\psi_{2}) \nonumber \\
                                              &+& v_{36}({\bf J};K_{c},L_{c})\cos 2\psi_{3} \nonumber \\
                                              &+& v_{231}({\bf J};K_{c},L_{c}) \cos(\psi_{1}-\psi_{2}-2\psi_{3}) \nonumber \\
                                              &+& v_{261}({\bf J};K_{c},L_{c})\cos(\psi_{1}-\psi_{2})
                                              \label{nonlinres}
\end{eqnarray}
In the above equations the actions  $(K_{c}=K+3/2,L_{c}=L+5/2,M_{c}=M+1/2)$ are conserved, being the classical analogs of the quantum polyads, and hence $H({\bf J},{\bm \psi})$ is ignorable in the conjugate angles $(\psi_{4},\psi_{5},\psi_{6})$.
The parameters $(C,{\bm \varpi},{\bm \alpha},{\bm \beta})$ of the reduced Hamiltonian (not given here) are
related to the original parameters via the canonical transformation and the reader is referred to a previous work for details\cite{jungtay}. It is important to note that the overlap of nonlinear resonances in Eq.~\ref{nonlinres} render the classical system nonintegrable with the possibility of extensive chaos in the classical phase space.  

\begin{center}
\begin{figure}[t]
\includegraphics[height=3.5in,width=3.0in]{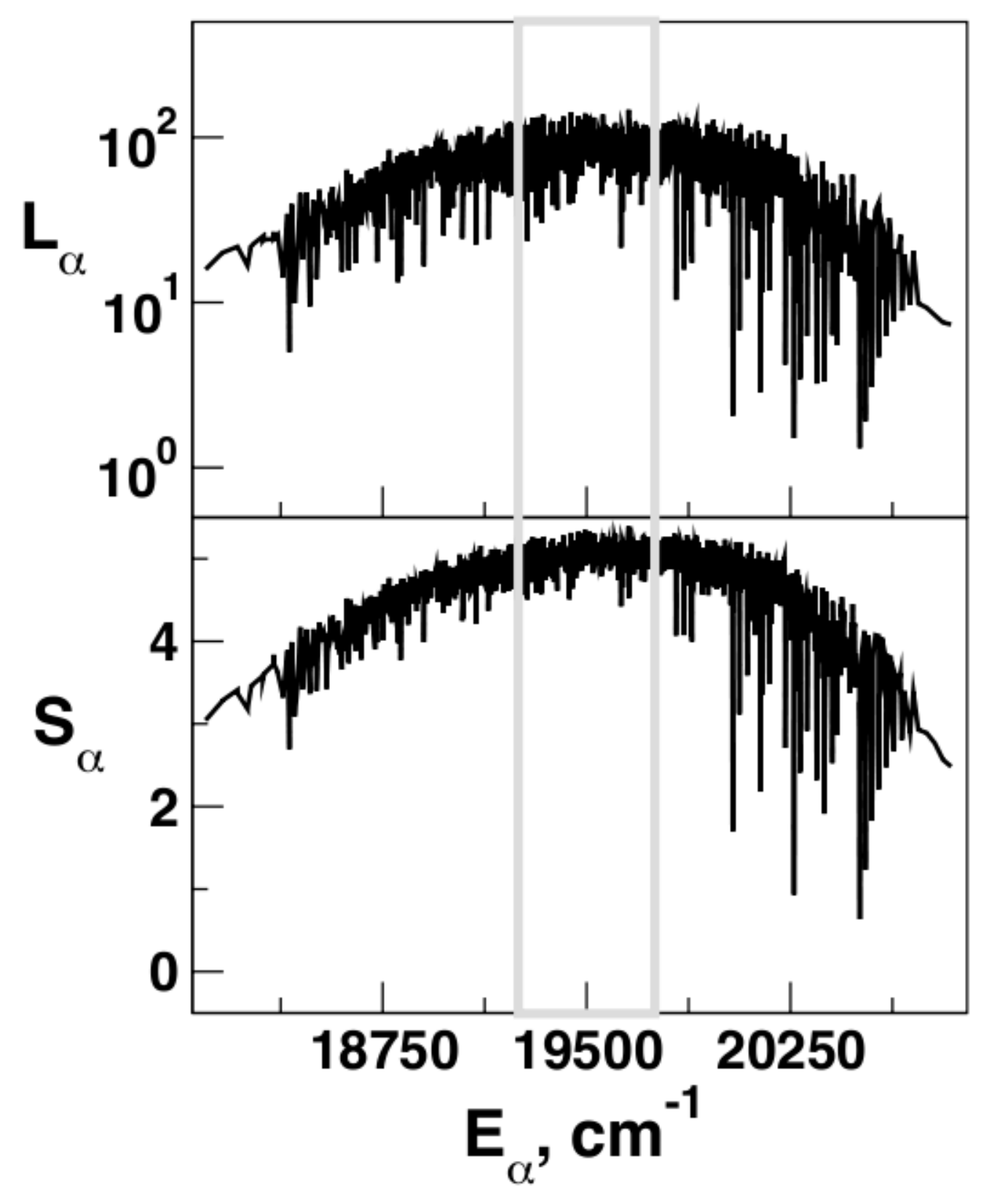}
\caption{Participation ratio (top panel) and Shannon entropy (bottom panel) for the eigenstates of SCCl$_{2}$ belonging to the polyad $(K,L,M)=(13,25,14)$. Both measures are computed in the zeroth-order number basis. The region indicated by gray rectangle comprises of the most delocalized eigenstates.}
\label{fig1}
\end{figure}
\end{center}

\section{Results and discussions}

We begin this section by computing two measures\cite{ipr} for eigenstate delocalization that are well known. The first quantity is the participation ratio
\begin{equation}
L_{\alpha} = \left(\sum_{\bf n}|\langle {\bf n}|\alpha \rangle|^{4}\right)^{-1} \equiv \left(\sum_{\bf n} p_{{\bf n} \alpha}^{2}\right)^{-1}
\label{ipr}
\end{equation}
which is a measure of the total number of basis states $\{|{\bf n}\rangle\}$ that participate in making up the eigenstate $|\alpha \rangle$. A second quantity is the information or Shannon entropy of an eigenstate given by the expression
\begin{equation}
S_{\alpha} = -\sum_{\bf n} p_{{\bf n}\alpha} \ln p_{{\bf n}\alpha}
\label{shanentropy}
\end{equation}
Note that the maximal\cite{ipr} $S_{\alpha} \approx \ln(0.48 N)$ for a gaussian orthogonal ensemble which has the value of $\sim6.5$ since in our case $N=1365$. It is also useful to observe that in Eq.~\ref{ethcor} if we choose $\hat{V} = |{\bf n}\rangle\langle {\bf n}|$ then $\bar{V} \equiv \sigma_{\bf n} = \sum_{\alpha}|C_{{\bf n}\alpha}|^{4}$ is the dilution factor associated with $|{\bf n}\rangle$ and assuming ETH one has $\sigma_{\bf n} = N_{\Delta E}^{-1}$. 

In Figure~\ref{fig1} the measures $L_{\alpha}$ and $S_{\alpha}$ are shown for all the eigenstates computed in the basis of the zeroth-order states {\it i.e.,} eigenstates of $H_{0}$. As expected both measures agree and indicate localized states at the low and high end of the spectrum. 
The middle part ~$(19250,19750)$ cm$^{-1}$ of the spectrum has states exhibiting large values for both the measures. However, note that $S_{\alpha}$ in particular does not become maximal even in this complicated energy range. Thus, although one expects that even at this high energy the eigenstates are not maximally delocalized, the origin of such partial localization are not apparent from the results.

\begin{center}
\begin{figure*}[t]
\includegraphics[height=4.0in,width=6.0in]{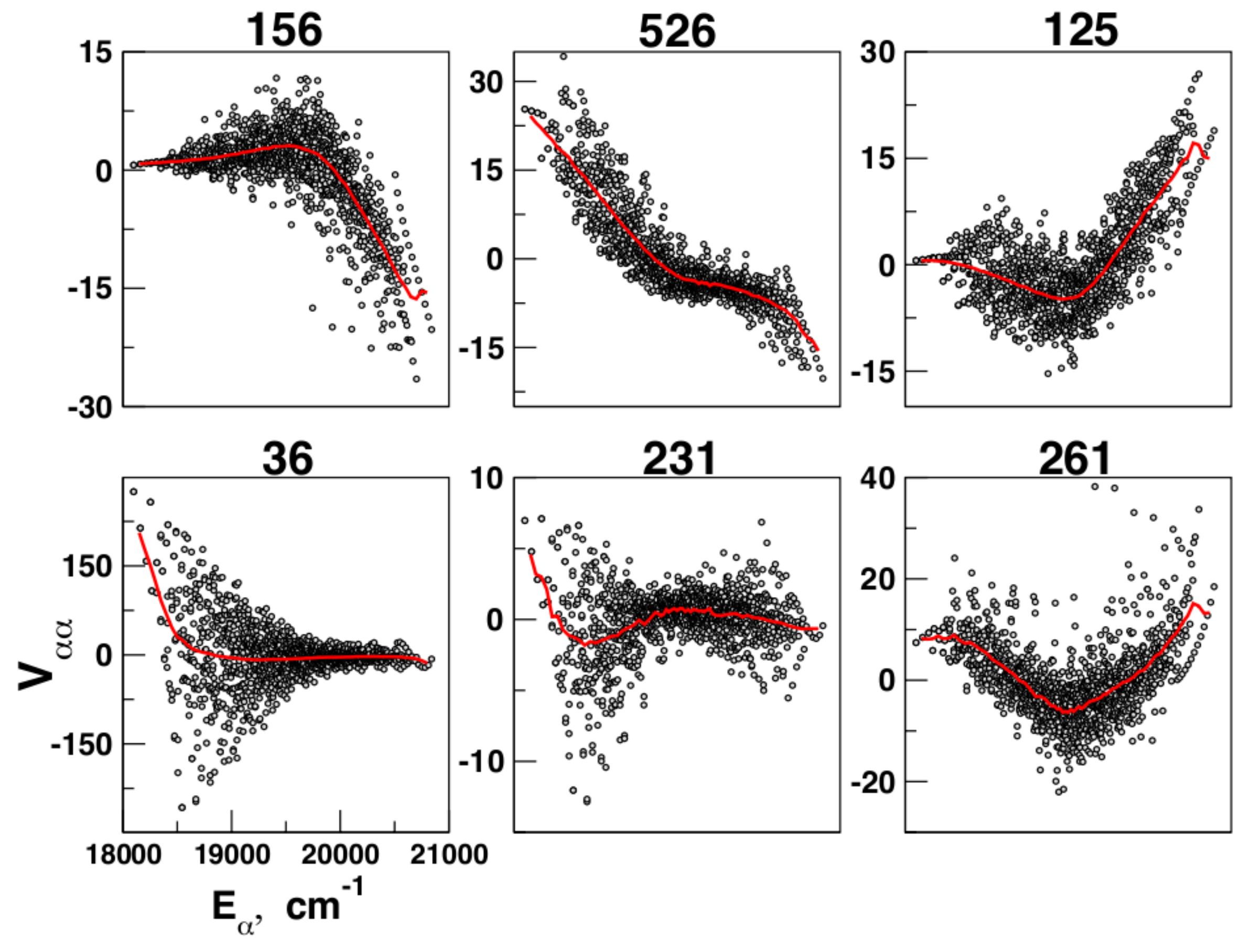}
\caption{Quantum eigenstate expectation values (circles) of the resonance operators (indicated on top of the panel) versus the classical microcanonical average (red line). The $x$-axis has the same scale in all plots and the $V_{\alpha \alpha}$ are dimensionless.}
\label{fig2}
\end{figure*}
\end{center}

 In order to gain more detailed insights we compute the eigenstate expectation values of the various resonance operators in Eq.~\ref{qham} and compare them to the  classical microcanonical average computed in the $({\bf J},{\bm \psi})$ reduced representation using the expression 
 \begin{equation}
\langle V \rangle_{\rm mc}(E) = \frac{\int V({\bf J},{\bm \psi}) \delta[H({\bf J},{\bm \psi})-E] d{\bf J}d{\bm \psi}}{\int \delta[H({\bf J},{\bm \psi})-E] d{\bf J}d{\bm \psi}}
\end{equation}
 The results, shown in Figure~\ref{fig2}, clearly show considerable fluctuations of the $V_{\alpha \alpha}$ about the $\langle V_{\rm mc} \rangle(E)$ and imply the existence of several localized eigenstates. Note that strongly localized states are seen, in agreement with Figure~\ref{fig1}, at both the low and high energy regions of the polyad. However, now it is possible to ascribe the low and high energy localized states with the $36$-resonance and a combination of the $156$ and $526$ anharmonic resonances respectively. Particularly striking is the $36$-resonance case at low energies exhibiting rather large quantum expectation values. In fact, following previous works\cite{dynassign_ks}, one can analyze the single integrable $36$-nonlinear resonance term in Eq.~\ref{nonlinres}
 \begin{equation}
 v_{36}({\bf J};K_{c},L_{c})\cos 2\psi_{3} \equiv 2 J_{3}(L_{1c}-K_{2c}-J_{3}) \cos 2\psi_{3}
 \end{equation}
 with $L_{1c} \equiv L_{c}-J_{1}$ and $K_{2c} \equiv K_{c}-J_{2}$ being constants. The analysis yields the extremal values for the quantum expectations as
 \begin{equation}
 \langle \alpha | a_{3}a_{3}a_{6}^{\dagger}a_{6}^{\dagger} + {\rm c.c}|\alpha \rangle = \pm\frac{1}{2}(L_{1c}-K_{2c})^{2}
 \label{ddmax}
 \end{equation}
 with $\pm$ corresponding to $\psi_{3}=0$ and $\psi_{3}=\pi/2$ respectively. Decent agreement of the above estimate with the computed values for the first few states shown in Figure~\ref{fig2} suggests that even around $\sim540$ THz above ground state the classical phase space of SCCl$_{2}$ has a large regular region capable of supporting several regular quantum states.  Using previously established techniques\cite{jungtay,dynassign_ks} it is possible to provide dynamical assignments for a number of states at the low and high end of the polyad. However, we do not attempt this here. Instead we focus on the $(19250,19750)$ cm$^{-1}$ part of the polyad shown in Figure~\ref{fig1} in order to check the extent of deviation from ETH in this region.
 
 \begin{center}
\begin{figure*}[t]
\includegraphics[height=4.0in,width=6.0in]{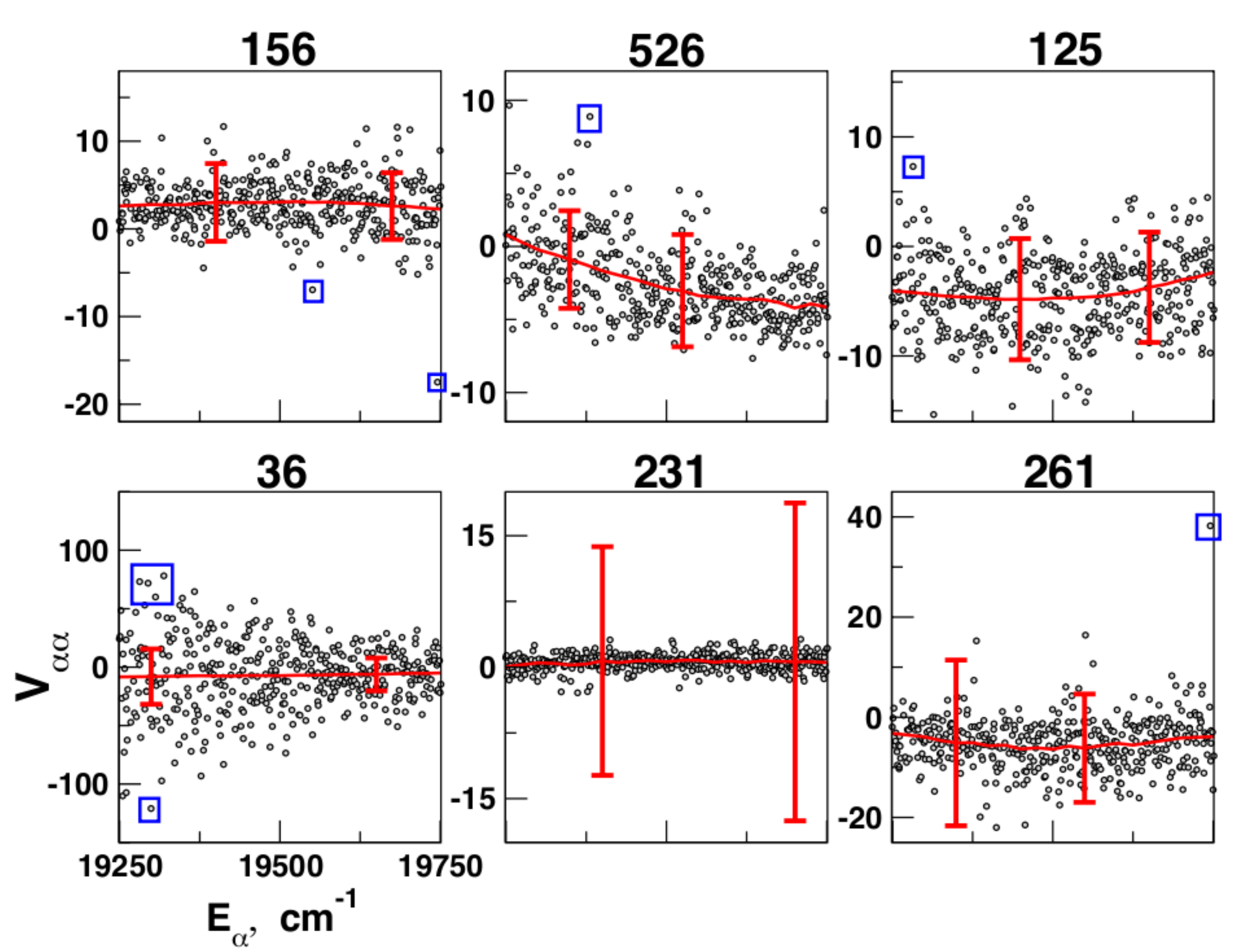}
\caption{Same as in Figure~\ref{fig2} but focusing on the complicated $(19250,19750)$ cm$^{-1}$ region of the polyad. The expected quantum fluctuations in $V_{\alpha \alpha}$ are also indicated at specific points. A few of the eigenstates lying outside the fluctuations, and hence partially localized, are highlighted by blue boxes.}
\label{fig3}
\end{figure*}
\end{center}

\subsection{Deviation from ergodicity}

The results in Figure~\ref{fig2} show the strong fluctuations about the classical microcanonical average suggesting deviations from ergodicity. However, as suggested by Pechukas\cite{pechukas}, the diagonal matrix elements $V_{\alpha \alpha}$ are expected to have quantum fluctuations which are of the same order of magnitude as the off-diagonal matrix elements $V_{\alpha \beta}$. It is therefore interesting to see if the deviations seen in the complicated $(19250,19750)$ cm$^{-1}$ region in Figure~\ref{fig2} are essentially the expected quantum fluctuations. 
As discussed by Feingold and Peres the computation of the quantum fluctuations in $V_{\alpha \alpha}$ is rather involved and related to the classical autocorrelation of the relevant observables\cite{eth_peresfeingold}. One approach is to model the distribution of $|V_{\alpha \beta}|^{2}$ by a Gaussian as in the earlier work\cite{eth_peresfeingold}. However,  in this work we take a more stringent constraint and associate the extent of quantum fluctuation of $V_{\alpha \alpha}$ with the largest off-diagonal matrix element $|V_{\alpha \beta}|$. Typically, the largest such element is found to be close to the diagonal. 

In Figure~\ref{fig3} we show the complicated region of the polyad with the expected quantum fluctuations in $V_{\alpha \alpha}$ at specific points. It is clear from Figure~\ref{fig3} that despite the near constant values for the various $V_{\alpha \alpha}$ there are several states (highlighted by boxes) that indeed deviate strongly from the microcanonical average. Interestingly, many of these states are not easily identifiable from Figure~\ref{fig1} since both $L_{\alpha}$ and $S_{\alpha}$ are not capable of identifying localization in the phase space. Hence, Figure~\ref{fig3} clearly establishes the power of the approach based on eigenstate expectation values. Note that different resonance operators reveal different localized eigenstates in a complementary fashion. For instance, a localized state with large negative expectation value for the $36$-resonance in Figure~\ref{fig3} is not easily identifiable from any of the other expectation values. Although not shown here, we have further confirmed  the predictions of Figure~\ref{fig3} by studying the semiclassical angle space representation\cite{jungtay} $\langle {\bm \psi}|\alpha \rangle \equiv \sum_{\bf n} c_{{\bf n}\alpha} \exp(i {\bf n} \cdot {\bm \psi})$ of the eigenstates.

The ETH result in Eq.~\ref{ethcor} essentially depends on the lack of correlation between the intensities $p_{{\bf n}\alpha}$ and the expectation values $V_{\alpha \alpha}$. Thus, are the results in Figure~\ref{fig3} contrary to the ETH expectations? In this context it is interesting to note that a sensitive measure for violations from ergodicity called as the intensity-velocity correlator was introduced earlier by Tomsovic\cite{cortom} and has been applied to several paradigmatic systems\cite{corapps}. This measure correlates the intensity $p_{z\alpha}\equiv |\langle z|\alpha \rangle|^{2}$ of some probe state $|z\rangle$ and the parametric evolution of the eigenvalues (level velocities) as
\begin{eqnarray}
C_{z}(\lambda) &=& \frac{1}{\sigma_{z}\sigma_{E}}\left \langle p_{z \alpha} \frac{\partial E_{\alpha}(\lambda)}{\partial \lambda} \right \rangle_{\alpha \in \Delta E} \nonumber \\
&\equiv& \left \langle \widetilde{p}_{z \alpha} \frac{\widetilde{\partial E_{\alpha}(\lambda)}}{\partial \lambda} \right \rangle_{\alpha \in \Delta E}
\label{intvelcor}
\end{eqnarray}
where  $\lambda$ is a Hamiltonian parameter of interest.  In the second line of Eq.~\ref{intvelcor} the quantities are scaled to unit variance and the level-velocities are zero centered to remove any net drift. The averaging above is done over a window $\Delta E$ consisting of $N_{\Delta E}$ states with $\sigma_{z}^{2}$ and $\sigma_{E}^{2}$ being the local variances of the intensities and level-velocities respectively. As defined Eq.~\ref{intvelcor} is expected to be large for eigenstates exhibiting common localization characteristics with levels moving in the same direction with changing parameter value. Moreover, one can show that in the random matrix theory (RMT) limit the correlator vanishes for every $|z\rangle$ up to fluctuations depending on the number of states within the energy window {\it i.e.,} $C_{z}(\lambda) = 0 \pm N_{\Delta E}^{-1/2}$. In this work, however, we will use the covariance instead of the correlator since $\sigma_{z}$ and $\sigma_{E}$ are also influenced by the classical phase space structures. Consequently, interpretation of $C_{z}(\lambda)$ needs to be done with some care\cite{corapps}. Moreover, using the covariance is closer to the spirit of Eq.~\ref{ethcor} and suffices for our work. In addition, note that an earlier work\cite{hellsofe} of Heller on the rate of exploration of phase space utilizes the strength function, which is the Fourier transform of the autocorrelation $\langle z(0)|z(t) \rangle$, as the central quantity. The $C_{z}(\lambda)$ is essentially the parametric response of the strength function\cite{cortom} and, as seen below, connects very naturally to the ETH perspective.

\begin{center}
\begin{figure}[t]
\includegraphics[height=3.0in,width=3.5in]{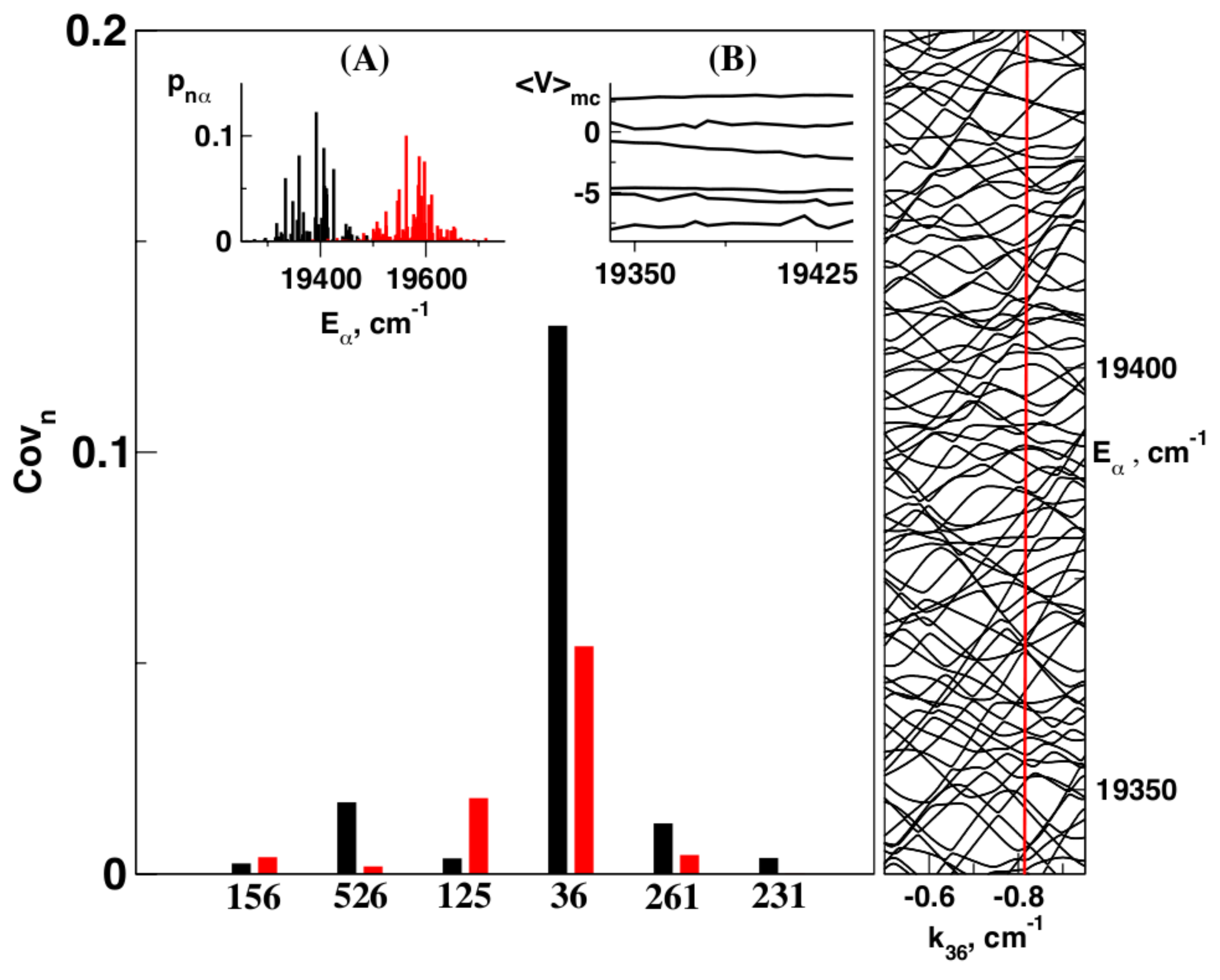}
\caption{Intensity-velocity covariance (cf. Eq.~\ref{correlate}) for two zeroth-order states $|{\bf n}\rangle = |5,8,15,14,0,0\rangle$ (black) and $|6,7,13,14,0,0\rangle$ (red). The resonant operators chosen to compute the covariances are indicated.  (A) Intensity spectra of the two states showing the fractionation due to resonances. (B) Classical microcanonical averages of the resonances vary very little over the energy window corresponding to the $|5,8,15,14,0,0\rangle$ state. Right panel shows the parametric evolution of the eigenvalues (level-velocities) upon varying the $36$-resonance strength. Actual resonance strength is indicated by a vertical red line.}
\label{fig4}
\end{figure}
\end{center}

To make contact with ETH we choose the parameter in Eq.~\ref{intvelcor} as one of the resonance strengths in Eq.~\ref{qham}, say $k_{36}$, and the initial state as the zeroth-order number state $|{\bf n}\rangle$. Then, using the Hellman-Feynman theorem, the covariance can be written down as
\begin{equation}
Cov_{\bf n}(k_{36}) =  \left \langle p_{{\bf n}\alpha} (V^{(36)}_{\alpha \alpha} - \bar{V}^{(36)}_{\Delta E}) \right \rangle_{\alpha \in \Delta E}
\label{correlate}
\end{equation}
where $\bar{V}^{(36)}_{\Delta E}$ is the window average of $V^{(36)}_{\alpha \alpha}$.  Similarly, one has measures for every anharmonic resonance in the Hamiltonian. Thus, as the expectations are zero-centered, ETH as in Eq.~\ref{ethcor} implies that $Cov_{\bf n}(k_{36}) = 0$ for every choice of $|{\bf n}\rangle$. On the contrary, nonzero values for $Cov_{\bf n}$ yield a quantitative measure of the deviations from ergodicity. 
We now choose two zeroth-order states $|5,8,15,14,0,0\rangle$ and $|6,7,13,14,0,0\rangle$ with energies $E_{\bf n}^{0} = 19388.3$ and $19574.7$ cm$^{-1}$ respectively in the range appropriate to Figure~\ref{fig3} and compute the intensity-velocity covariance for each resonance operator.  The energy windows  are centered at their respective $E_{\bf n}^{0}$ with widths $\Delta E = 2 \delta E_{\bf n}^{0}$ determined by the energy uncertainty $\delta E_{\bf n}^{0}$ ($\approx 50$ cm$^{-1}$ for both states). Within the selected window there are about $80$ eigenstates and the respective classical microcanonical expectations are nearly constant (cf. Figure~\ref{fig3}(B)). Hence, the classical phase space is not expected to change much over the averaging window. The results are shown in Figure~\ref{fig4} and clearly indicate deviations from ergodicity. As expected, the $36$-resonance plays an important role for both the states. However, localization due to the $526$ and the $125$ resonances are also observed.  The dominance of the $36$-resonance can be understood from the parametric evolution of the eigenvalues shown in Figure~\ref{fig4} with varying resonance strength. Several states exhibiting linear parametric motion (``solitonic" states\cite{pechukas,gaspard}) can be seen amidst a sea of avoided crossings. Such solitonic states are the robust localized states, also seen exhibiting large fluctuations in Figure~\ref{fig3}, resulting in the strong deviations from ergodicity as measured by the covariance.

A final remark is in order at this stage regarding the choice of the two specific zeroth-order states chosen in Figure~\ref{fig4}.  In their recent work\cite{chowgrueb1,chowgrueb2} Chowdary and Gruebele associated the sharp SEP features ($\sim600$ THz)  with zeroth-order states of the form $|n_{1},n_{2},n_{3},n_{4},0,0\rangle$ with $n_{3}$ being typically rather small ($\approx 0$ as seen in a later work\cite{chowgrueb2}) in comparison to the other mode occupancies.  Although the states chosen here have the same form, the $n_{3}$ occupancies being large makes them strongly susceptible to the $36$-resonance and hence fragmented in the SEP spectra. However, such states are fragmented much less compared to the states at similar energies but with occupancy in the $n_{5}$ and $n_{6}$ modes. For example, the zeroth order state $|3,6,9,14,4,6\rangle$ with $E_{\bf n}^{0} \approx 19389.1$ cm$^{-1}$ is diluted by nearly an order of magnitude more as compared to the state $|5,8,15,14,0,0\rangle$. Thus, the results in Figure~\ref{fig4} showing considerable localization at these high energies are compatible with the findings of Chowdary and Gruebele\cite{chowgrueb1,chowgrueb2}. 

\section{Conclusions}
In this work we have shown that several eigenstates of SCCl$_{2}$ exhibit localization even near the dissociation threshold. Such `nonthermalized' states are identified using sensitive and quantitative measures for deviations from ergodicity. Consequently, the dynamics at these energies should exhibit deviations from RRKM predictions. However, as pointed out in an earlier work\cite{chowgrueb2}, observing the deviations from RRKM depends on whether the optically accessible states have sufficient overlaps with the partially localized eigenstates or not. In other words, experimentally prepared initial states which have significant overlaps with partially localized eigenstates are bound to have nonstatistical dynamics since the eigenstates encode the infinite time energy flow dynamics of the system. Although the converse statement {\it i.e.,} validity of ETH implies RRKM is yet to be established, the important studies\cite{nordhrice} by Nordholm and Rice does lead us to believe that it is reasonable to expect so. It is also relevant to note that the present work is an example of a system wherein the integrability breaking perturbations do have specific selection rules and hence testing the ETH in a more complex situation\cite{ethhom}.

The techniques used here are very general and not limited by the dimensionality of the system.  The intensity-velocity correlator used in this study is not only capable of identifying localized states but also singles out the relevant perturbations that lead to localization. Note that the eigenstate expectation values play a key role in both ETH and the matrix fluctuation-dissipation (MFD) approach\cite{mfd} of Gruebele.  Hence, a better understanding of the connection between the two should be useful in gaining fundamental classical-quantum correspondence insights into the dynamics and  control of IVR.  Finally, it would be of some interest to determine the exact nature of the relation between ETH on one hand, and the Logan-Wolynes transition criterion\cite{lwjcp}, and the Leitner-Wolynes quantum ergodicity criterion\cite{leitwoly96}, which includes the dynamical tunneling effects, on the other hand.

\pagebreak

\end{document}